\documentclass[preprint2]{aastex}
\usepackage{graphicx}
\usepackage{color}

\begin{document}

\title{Prospects of SETI by small size optical telescopes}

\author{Z.N. Osmanov$^{1,2}$ \& V.I. Berezhiani$^{1,3}$}
\affil{$^1$School of Physics, Free University of Tbilisi, 0183, Tbilisi,
Georgia}
\affil{$^2$E. Kharadze Georgian National Astrophysical Observatory, Abastumani, 0301, Georgia}
\affil{$^3$Andronikashvili Institute of Physics (TSU), Tbilisi 0177, Georgia}

\begin{abstract}
In the present manuscript we consider the possibility of SETI by the small size (with diameters less than $1$ m) optical telescopes. Calculations are performed for typical parameters of the mentioned type of  telescopes. In particular, we show that the techno-signatures of Type-2.x and Type-3.x civilizations might be detected. It is demonstrated that it is possible to detect as hot megastructures (up to $4000$ K) built around main sequence stars and pulsars as well as von Neumann extraterrestrial probes.

\end{abstract}


\section{Introduction}

In the present paper we consider the possibility of search for extraterrestrial intelligence (SETI) by small size (with diameters up to $1$ m) optical telescopes and we discuss how prospective this project is. It is quite clear that SETI means the search for peculiarities in the detected emission and in the present paper we focus on the search for technosignatures of advanced alien civilizations. Today many relatively old observatories gradually become either passive or transform into museums. But they can be used in the mentioned search, when large telescopes cannot spend much time on SETI projects.

In 1960 a very original idea of the search for technosignatures has been proposed by Freeman Dyson \citep{dyson}. The author has assumed that the alien technological society is advanced enough to consume the whole energy of their host star, belonging to Type-II civilization in Kardashev's classification. According to this ranking Type-I society is consuming the power of solar radiation incident on Earth and Type-III civilization utilizes almost the total power of the host galaxy \citep{kardashev}, but since more appropriate designation of technological level is fractional \citep{Cirkovic}, henceforth we use Type-1.x; Type-2.x and Type-3.x. \cite{dyson} has suggested that to consume the whole energy of the host star a civilization has to build a gigantic spherical megastructure - Dyson sphere (DS) - around a host star in the habitable zone. As a result, this megastructure will inevitably emit in the infrared spectral band and thus it will be potentially detectable.

In the end of the last century and the beginning of the 21st century several attempts have been performed to search for DSs  \citep{jugaku,slish,timofeev,carrigan} and despite the fact that some of the observational features have have been identified as potential candidates it has been emphasised that further study is necessary. It is worth noting that silence of the universe (Fermi paradox) means that the search for the extraterrestrial intelligence is not that trivial and therefore, one has to examine as many channels as possible. 

By \cite{paper1,paper2} the possibility of building ring-like megastructures located in the habitable zone of pulsars has been considered and the possibility of their detection in the infrared spectrum by modern facilities was studied. On the other hand, monitoring the sky only in the infrared spectrum significantly restricts the methods of search. To extend a spectral area of the search \cite{paper3} have considered the possibility of hot DSs (considering Graphen as an example) with temperatures up to $4000$ K and thus visible in the optical spectral band. It has been shown that in case of hot megastructures an amount of material necessary to build a DS will be much less than for the megastuctures located in the habitable zone. Another interesting feature the observational signatures of megastructures will be characterised is the spectral variability. \cite{paper5} and \cite{fermi} have examined ring-like megastructures around stars and pulsars and it was found that the hot rings oscillate nearby equilibrium positions leading (via the Doppler shift) to the spectral variability. The corresponding time-scales vary from several minutes (M-type stars) up to years (O-type stars). In case of pulsars the variability period is of the order of several days.

Another class of extraterrestrial engineering technosignatures of which we intend to consider in the context of their search by small size telescopes is the von Neumann self-reproducing probes. Recently extending an idea of von Neumann  \citep{neum}, extraterrestrial self-reproducing probes have been considered \citep{probe1,probe2} and it has been shown that by collecting material in the interstellar media for replication, the probes will be visible in a broad electromagnetic spectral band and as it will be shown, also in the optical spectra.

The aim of the present paper is to consider the possibility of detection of technosignatures produced by the megastructures and von Neumann probes by small size optical telescopes. And as an example the AbAO telescope AZT-14 with the aperture $48$ cm \citep{azt} is taken to show how useful smaller aperture telescopes are.

The paper is organized in the following way: in Sec. 2, we briefly outline the general properties of technosignatures (megastructures and von Neumann probes), discussing the possibility of their detection and in Sec. 3 we summarise obtained results.

\section[]{Main Consideration}

  In this section we intend to outline the results presented in \citep{paper5,probe2,probe1,fermi} to consider the observational features of technosignatures in the visible spectra and analyse the possibility of their detection by a typical small size telescope.  
\subsection[]{Hot megastructures}

As it has been studied in detail by \citep{wright} it is unrealistic to construct a monolithic spherical megastructure, because no material can be stable against internal gravitational stresses, implying that DSs should be composed of concentric rings. Then, one can show that the radius of the ring is given by \citep{paper5,fermi}

$$R=
\left(\frac{L}{8\pi\sigma T^4}\right)^{1/2}\simeq $$
\begin{equation}
\label{R} 
\simeq 6.9\times 10^{-3}\times 
\left(\frac{L}{L_{\odot}}\right)^{1/2}\times
\left(\frac{4000K}{T}\right)^2 AU,
\end{equation}
where $L$ is the bolometric luminosity of the star, normalized by the Solar luminosity, $L_{\odot}\approx 3.83\times 10^{33}$ergs s$^{-1}$, $\sigma\approx 5.67\times 10^{-5}$erg/(cm$^2$K$^4$) denotes the
Stefan-Boltzmann's constant and $T$ is the temperature of the megastructure. From Eq. (\ref{R}) it is clear that $R$ is much less than the typical radii of the DS located in the habitable zone ($\sim 1$AU). From the Wien's law it follows that the spectral radiance peaks at the wavelength
\begin{equation}
\label{lamb} 
\lambda_{peak} = \frac{2898 \mu m\;K}{T} \simeq 725\times 
\frac{4000 K}{T}\; nm,
\end{equation}
which as is clear from Eq. (\ref{lamb}) is of the order of $725 $nm for $T = 4000$ K. Angular resolution of AZT-14 (with a diameter $D=48$cm) for the mentioned maximum temperature $\varphi = 1.22\;\lambda_{peak}/D\approx 1.8\times 10^{-6}$ rad is not enough to resolve the structure of the ring from the cosmic distances. 

One the other hand, one can straightforwardly check that if a maximum value of the magnitude detectable by the telescope is $m$, then the maximum distance where the megastructure with radius $R$ is still visible for the corresponding facility is given by
\begin{equation}
\label{Dm} 
D_{m} \simeq R\times10^{\frac{m}{5}}\times \sqrt{\frac{B_{\nu}(T)}{F_0}},
\end{equation}
 where 
 \begin{equation}
\label{Bn} 
B_{\nu} = \frac{2h\nu^3}{c^2}\frac{1}{e^{h\nu/(kT)}-1}
\end{equation}
is the spectral radiance of the black body emission, $h$ represents the Planck's constant, $\nu$ is the frequency, $F_0$ is the spectral flux at $m = 0$ for the given temperature $c$ is the speed of light and $k$ represents the Boltzmann constant. As an example, we consider the highest possible temperature ($T = 4000$ K), then $F_0\approx 2.55\times 10^{-20}$erg s$^{-1}$cm$^{-2}$Hz$^{-1}$ (for $\lambda_{peak} = 725$nm). 

In photometric observations the limiting magnitudes are much higher compared to the limiting magnitudes in spectrometric observations. We make calculations for moderate values of magnitudes. In particular, if we assume that the achievable magnitude in spectrometric measurements is at least of the order of $5$ for the integration time $\sim 1$ min \footnote{From private communication with Prof. N. Kochiashvili}, then, one can straightforwardly show that for the integration times of the order of $15$ min and $40$ min the maximum values of magnitudes will be respectively $8$ and $9$. 
Equation (\ref{Dm}) leads respectively to the following maximum distances $D_{8}\approx 100$ pc, $D_{9}\approx 160$ pc. By taking into account that the number density of G-type stars (solar type stars) in the solar neighbourhood equals $\sim 3.2\times 10^{-3}$ pc$^{-3}$ \citep{density} it is straightforward to show that one can monitor $1.3\times 10^4$ stars (if $m = 8$), or $5.5\times 10^4$ stars (if $m = 9$). The similar analysis performed for other  values of temperatures also leads to a large number of solar-type stars. Generally speaking, it is worth noting that the problem of identification of DSs is complex and in order to distinguish a Dyson megastructure from a star one should use several filters.

It is worth noting that if a DS is not complete multi temperature black body radiation patterns (one from the star and the rest from rings generally having different radii) will be detectable.

Apart from detecting flux there is another interesting feature. As it has been shown by \cite{fermi,paper5} the rings will be stable against out-off plane motions oscillating nearby the equilibrium state and it was found that the period of oscillations writes as
 $$P = 2\pi\sqrt{\frac{R^3}{GM}}\simeq$$
 \begin{equation}
\label{P} 
\simeq 2.5\times\left(\frac{M_{\odot}}{M}\right)^{1/2}\times\left(\frac{L}{L_{\odot}}\right)^{3/4}\times\left(\frac{4000\;K}{T}\right)^3\; hours,
\end{equation}
where $M_{\odot}\simeq 2\times 10^{33}$ g represents the solar mass and $G$ is the gravitational constant. 
It is evident that the time-scale of oscillations for hot megastructures is several hours, which potentially might be detectable because the periodic motion of the megastructures will cause the spectral variability of the emission pattern. In particular, from the Doppler shift one can straightforwardly derive that the maximum value of the wavelength difference is given by
 \begin{equation}
\label{dlamb} 
\Delta\lambda\simeq\lambda\frac{2\upsilon_m\cos\theta}{c},
\end{equation}
where $\lambda$ is the wavelength of emission in the rest frame of the megastructure, $\upsilon_m = 2\pi A/P$ is the velocity amplitude of oscillations, $A$ is the spacial amplitude of oscillations and $\theta$ represents an angle of velocity direction measured in the observer's frame of reference. There are very high precision modern spectrometers with the resolving power $RP = \lambda/\Delta\lambda=50000$,  \footnote{http://www.ltb-berlin.de/}. On the other hand, it s worth noting that for not making overestimation in the precision of measurements, one should estimate the maximum values of $RP$ limited by telescopes. In particular, if the minimum detectable dimensionless flux difference for a given telescope is $\Delta F/F = 10^{-\frac{2m}{5}}$, then, by equating it with the corresponding value for the black body emission of a megastructure, $\Delta B_{\lambda}/B_{\lambda}$ one obtains the following expression for the resolving power
\begin{equation}
\label{RP} 
RP_{max}\simeq10^{{\frac{2m}{5}}}\times\left(5-\frac{hc}{\lambda kT}\frac{e^{hc/(\lambda kT)}}{e^{hc/(\lambda kT)}-1}\right),
\end{equation}
which for the aforementioned magnitudes $m = \{8, 9\}$ leads to values, $RP_{max} = \left(5600, 14200\right)$.

One can show that the spectral variations might be detected if the normalised spacial amplitude, $X = A/R$, satisfies the following condition
$$X\geq \frac{c}{2RP\;T\cos\theta}\times\left(\frac{L}{8\pi\sigma G^2M^2}\right)^{1/4}\simeq $$
$$\simeq 0.07\times \frac{1}{\cos\theta}\times\frac{6000}{RP}\times$$
\begin{equation}
\label{cond1} 
\times\frac{4000 K}{T}\times\left(\frac{M_{\odot}}{M}\right)^{1/2}\times\left(\frac{L}{L_{\odot}}\right)^{1/4},
\end{equation}
where the resolving power is normalised by $6000$. For $RP = 5600$ $X\simeq 0.08$ and for $RP = 14200$ the corresponding value is even less: $0.03$. Since $X$ should be small the search for hot megastructures built around main sequence stars by small size optical telescopes seems to be realistic.

The similar calculations can be performed for megastructures built around pulsars. In particular, as it was derived by \cite{fermi} the ring's radius is given by
$$r\simeq
\left(\frac{L_p}{4\pi\sigma\beta T^4}\right)^{1/2}\simeq $$
\begin{equation}
\label{R1} 
\simeq 1.1\times 10^{-4}\times 
\left(\frac{L_p}{10^{30}\; erg/s}\right)^{1/2}\times
\left(\frac{4000K}{T}\right)^2 AU,
\end{equation}
where $L_p$ is the normal period (of the order of $1$ sec) pulsars' bolometric luminosity, normalised by the typical value and $\beta\simeq 32^0$ \citep{ruder} is an opening angle of the pulsar's emission cone (in Eq. (\ref{R1}) it is written in radians). By taking into account the obtained expression, one can straightforwardly show that the maximum distance where a ring-like megastructure around a pulsar is visible
\begin{equation}
\label{Dm1} 
d_{m} \simeq r\times10^{\frac{m}{5}}\times \sqrt{\frac{\beta B_{\nu}(T)}{F_0}}
\end{equation}
for the aforementioned values with $m = \{8,9\}$ leads to $d_8\simeq 280$ pc and $d_9\simeq 440$ pc respectively. In Eq. (\ref{Dm1}) $\beta$ is substituted in radians. By combining $d_{8,9}$ with the pulsars' surface number density of distribution in the galactic plane, $N_p\simeq 520$ kpc$^{-2}$ \citep{manchester}, one obtains that approximately $N\simeq \pi d_{8}^2N_p \simeq 130$ (for magnitude $8$) and $N\simeq \pi d_{9}^2N_p \simeq 320$ (for magnitude $9$) pulsars can be monitored. On the other hand, it is clear that pulsars are detected when their beams periodically radiate toward Earth and many of them remain undetected by means of the pulsed emission. But in the framework of our approach they still can be monitored, since the observations should be conducted for the second stage optical emission of megastructures.

After taking into account the typical value of pulsar's mass $M_p\simeq 1.5 \times M_{\odot}$, from Eq. (\ref{P}) one finds that a variability timescale is of the order of $\sim 1$ min (although it might be higher by one order of magnitude depending on luminosity). Likewise the previous case one can obtain the condition for the dimensionless amplitude of oscillations $\chi = A/r$
$$\chi\geq \frac{c}{2RP\;T\cos\theta}\times\left(\frac{L_p}{8\pi\sigma G^2M_p^2}\right)^{1/4}\simeq $$
$$\simeq 0.09\times \frac{1}{\cos\theta}\times\frac{6000}{RP}\times$$
\begin{equation}
\label{cond2} 
\times\frac{4000 K}{T}\times\left(\frac{1.5\times M_{\odot}}{M_p}\right)^{1/2}\times\left(\frac{L_p}{10^{30}\; erg/s}\right)^{1/4}
\end{equation}
when the spectral variability might be detected by a high resolution spectrometer. In case of $RP = 17000$, the minimum value becomes $0.03$. 

\subsection[]{von Neumann probes}

In this subsection we refer to the recent works by \cite{probe2,probe1} to examine the emission signatures of extraterrestrial von Neumann probes in the context of their detectability.

\subsubsection[]{Self replication}


First of all, we would like to discuss a process of self-reproduction. In due course of time, the probes will collect material for replication. Then, from the mass injection rate $\dot{M} = \pi\beta r_0^2m_0nc$, one can straightforwardly show that \citep{probe2} the replication time-scale, $\tau = M/\dot{M}$, of a spherical probe moving in a hydrogen cloud is of the order of 
$$\tau = \frac{4\xi}{3\beta}\times\frac{\rho}{m_0n}\times\frac{r_0}{c}\simeq 0.74\times \frac{\xi}{0.1}\times\frac{0.05}{\beta}\times$$
\begin{equation}
\label{tau} 
\;\;\;\;\times\frac{m_p}{m_0}\times \frac{\rho}{0.4g\;cm^{-3}}\times\frac{10^4cm^{-3}}{n}\times\frac{r_0}{0.1 mm} \;yrs,
\end{equation}
where $\xi < 1$ denotes the fraction of the probe's total volume filled with the material, $\rho$ is its density normalized by the density of Graphene, $\beta = \upsilon/c$, denotes the probe's dimensionless velocity, $m_0$ is the mass of encountering molecules, $n$ is their number density in a cloud and $r_0$ is the probe's radius. In \citep{probe2,probe1} it was shown that an optimal size of the probes to explore a certain area of space strongly depends of its parameters (composition, density and size).  For a spherical hydrogen cloud with the number density $10^4 cm^{-3}$ and radius $1$ pc the size equals $0.013$ mm and for Type-3.x civilization exploring the whole galaxy with the diameter $50000$ pc and the number density $1 cm^{-3}$ one obtains $0.25$ mm. Collecting material will inevitably lead to a drag force with the corresponding power $P_{_{Drag}}\simeq\upsilon^2dM/dt$ \citep{probe2}, where $dM/dt$ denotes the mass rate of the collecting material. On the other hand, if propulsion mechanism is provided by thermonuclear fusion and for that reason a certain fraction $\alpha$ of the total mass is used, the power will be given by $P\simeq\alpha\;\epsilon\;c^2dM/dt$, where $\epsilon\simeq 0.0035$ is an approximate value of the fraction of total rest energy that might be utilised in thermonuclear reactions \citep{carroll}. After equating both quantities, one arrives at the conclusion $\beta\simeq\sqrt{\alpha\epsilon}$, leading to the maximum possible velocity (for $\alpha = 1$) $\beta_m\simeq\sqrt{\epsilon}\simeq 0.06$. In this case the total material is used for propulsion with no possibility of replication. One can straightforwardly check that the mass fraction used for replication then writes as $1-\alpha\simeq 1-\beta^2/\beta_{max}^2\simeq 0.3$. Consequently the time-scale of replication becomes $\tau/(1-\alpha)\simeq 2.5$ yr. The thermonuclear process itself can be provided by collision of hydrogen atoms inside a probe. 

It is worth noting that exploring different types of clouds with different composition, the replication time-scale will be different. In particular, one can straightforwardly check that for giant molecular clouds with $CO$ in the dense cores, where the number density might range from $10^4$ cm$^{-3}$ to $10^6$ cm$^{-3}$, the reproduction time-scale will be of the order of $150$yrs (for $10^4$ cm$^{-3}$)  and $1.5$yrs (for $10^6$ cm$^{-3}$). But in the manuscript we focus on calculations for hydrogen, since it is the most abundant element in the universe. One should also emphasise that at the end of exploration the overall exponential growth must be terminated because at the edge of the nebula, newly produced probes must spend more time to travel to new materials. Apart form that, it is also possible that one part of the swarm might finish before another. 

From the classical point of view the thermonuclear reactions start when the distance between nuclei become of the order of $(10^{-15}-10^{-14})$ cm, but due to the wave-like nature of nuclei, the mentioned extremely small distances are not necessary, it can be guaranteed by quantum tunnelling, leading to a sufficiently high effective temperature enough for starting thermo-nuclear reactions \citep{carroll}.

\subsubsection[]{Spectral characteristics}

Emission spectral feature of the probes might be quite complex. We consider only a simplified picture to make estimates. In particular, it is assumed that if the atoms are captured by the probes, from the point of view of a laboratory observer the particles which initially have been almost at rest now are accelerated, with almost constant acceleration. On the other hand, the process of capturing might be provided by very strong magnetic field and although the spectra (of the synchrotron mechanism) will depend on morphology of the magnetic field, the general picture will be the same. It is also clear that the capturing process will lead to thermal emission, but for identification of a swarm of probes some distinguished feature has to be detected, therefore we focus on an emission pattern generated by acceleration process of charges. By \cite{probe2} it was shown that the process of collecting the space material (hydrogen molecules) leads to efficient emission of a swarm of von Neumann probes characterised by the following spectral flux
\begin{equation}
\label{flux} 
\frac{dW}{dtdf} = 2^{t/\tau}n N_0\frac{4\pi e^2r_0^2\beta^3}{3}\times\left(\frac{f_0}{f}\right)^2\sin^2\left(\frac{f}{f_0}\right),
\end{equation}
where $N_0$ is the initial number of probes, $f$ is the emission frequency normalized by the factor $f_0$, given by \citep{probe1}
\begin{equation}
\label{freq} 
f_0\simeq\frac{\beta c}{2\pi r\kappa}\simeq  2.4\times 10^{11}\times\frac{0.1}{\kappa}\times\frac{\beta}{0.05}\times\frac{0.1 mm}{r}\; Hz,
\end{equation}
where $\kappa$ is the fraction of the probe's size, where the hydrogen particles are captured and throughout the paper we assume $\kappa = 0.1$. It is evident that the radiation pattern covers the whole range of electromagnetic spectrum but with different efficiency. For optical telescopes one should focus on the visible light, with frequencies in the following interval $(4-8)\times 10^{14}$ Hz. One of the interesting features of Eq. (\ref{flux}) is that the spectral flux diminishes at discrete values of frequencies: $f_k = \pi k f_0$, which might be considered as one of the significant fingerprints of existence of von Neumann probes. Considering Type-2.x civilisation with $N_0 =100$ one can straightforwardly show from Eq. (\ref{flux}) that at the end  exploration of the spherical nebula with radius $1$ pc, the spectral flux corresponding to the same wavelength as in the previous cases, $\lambda = 725$ nm, is of the order of $3\times 10^{18}\; erg\ s^{-1}\; Hz^{-1}$ ($\beta = 0.05$), which when substituted into Eq. (\ref{Dm}) instead of $B_{\nu}(T)$ leads to the maximum distances of detection of the order of $40$ pc if $m = 8$ and $60$ pc for $m = 9$. In the similar way, for Type-3.x civilizations, for the same velocity one obtains $1.7$ Mpc. An important fingerprint of von Neumann self-replicators is the exponential growth of spectral flux (see Eq. \ref{flux}), characterised by the time-scale, $\tau^{\star} = \tau/((1-\alpha)\ln 2)$, which is of the order of $3.6$ yr, indicating that the process of flaring of a certain region of space might be observed in a realistic time-scale. One should also emphasise that the net effect of the glow will have a diffusive character, potentially affecting the sensitivity.

\section{Conclusion}

In the present manuscript we show that small size optical telescopes can detect the flux from the hot ring-like megastructures built around main sequence stars and pulsars. It is shown that to search for the candidates the maximum distance will be of the order of $160$ pc for stellar megastructures and $440$ pc for pulsar megastructures.

The possible extension of the search methods has been examined by considering the variability of megastructures and it was found that the corresponding timescales are of the order of several hours, which might be detected by using high performance spectrographs with optimal resolving powers $R_p = 6000$ (for $m = 8$)  and $R_p = 17000$ (for $m = 9$).

We also considered the possibility of detection of extraterrestrial von Neumann probes and it has been found that by using a small size optical telescope the detection of the mentioned objects is quite realistic on distances: $60$ pc for Type-2.x civilizations and $1.7$ Mpc for Type-3.x civilizations.

The mentioned analysis indicates how prospective SETI is at ground based optical small size telescopes.

\section{Acknowledgements}
The authors would like to thank Prof. G. Javakhishvili, Prof. N. Kochiashvili and an anonymous referee for fruitful suggestions and comments.


\begin{thebibliography}{99}

\bibitem[Bovy(2017)]{density} Bovy, J. 2017, MNRAS, 470, 1360
\bibitem[Carroll \& Ostlie(2010)]{carroll} Bradley W. Carroll \& Dale A. Ostlie, An Introduction to Modern Astrophysics and Cosmology, Cambridge University Press, Cambridge, UK, 2010
\bibitem[Carrigan(2009)]{carrigan} Carrigan, R. A., 2009, ApJ, 698, 2075 
\bibitem[Cirkovic(2015)]{Cirkovic} Cirkovic, M.M. 2015, SerAJ, 191, 6
\bibitem[Dyson(1960)]{dyson} Dyson, F., 1960, Science, 131, 1667  
\bibitem[Jugaku \& Nishimura(2004)]{jugaku} Jugaku, J. \& Nishimura, S., 2004,
in Proc. IAU Symp. 213, Bioastronomy 2002: Life Among the Stars, ed.
R. Norris \& F. Stootman (San Francisco, CA: ASP), 437
\bibitem[Kardashev(1964)]{kardashev} Kardashev, N. S., 1964, Sov. Astr., 8, 217  
\bibitem[Kumsiashvili  \& Kraicheva(1982)]{azt} Kumsiashvili, M. I.\& Kraicheva, Z., Proceedings of the Sixty-ninth Colloquium, Bamberg, West Germany, August 31-September 3, 1981. (A82-48076 24-90) Dordrecht, D. Reidel Publishing Co., 1982, p. 467-472 
\bibitem[Manchester et al.(2005)]{manchester} Manchester, R.N. et al. The Australia Telescope National Facility Pulsar Catalogue, AJ, 2005, 129, 1993
\bibitem[von Neumann(1966)]{neum} Von Neumann, J., 1966, Theory of Self-Reproducing Automata 
(University of Illinois Press, Urbana and London)
\bibitem[Osmanov(2021)]{fermi} Osmanov, Z.N., 2021, IJAAE, 6, 049
\bibitem[Osmanov(2020a)]{probe2} Osmanov, Z., 2020a, JBIS, 73, 7
\bibitem[Osmanov(2020b)]{probe1} Osmanov, Z., 2020b, IJAsB, 19, 220
\bibitem[Osmanov(2017)]{paper2} Osmanov, Z., 2018, IJAsB, 17, 112
\bibitem[Osmanov(2016)]{paper1} Osmanov, Z., 2016, IJAsB, 15, 127
\bibitem[Osmanov and Berezhiani(2019)]{paper5} Osmanov, Z. and Berezhiani, V.I., 2019, JBIS, 72, 254
\bibitem[Osmanov and Berezhiani(2018)]{paper3} Osmanov, Z. and Berezhiani, V.I., 2018, IJAsB, 17, 356
\bibitem[Ruderman \& Sutherland(1975)]{ruder} Ruderman M.A. \& Sutherland P.G., 1975, ApJ, 196, 51
\bibitem[Papagiannis(1985)]{slish} Papagiannis, M. D., 1985, The Search for
Extraterrestrial Life: Recent Developments, ed. (Boston, MA: Reidel Pub. Co.), 315
\bibitem[Timofeev et al.(2000)]{timofeev} Timofeev, M. Y., Kardashev, N. S. \& Promyslov, V. G., 
2000, Acta Astronautica J., 46, 655
\bibitem[Wright(2020)]{wright} Wright, J., 2020, Serb. Astron. J., 200, 1


\end{thebibliography}
\end{document}